\documentstyle[aps,prl,graphicx]{revtex}

\input psfig.sty
\draft
\begin{document}
\twocolumn 
\wideabs{  
\title{Observation of anomalous spin-state segregation in a trapped ultra-cold vapor}
\author{H.~J. Lewandowski, D.~M. Harber, D.~L. Whitaker \cite {williams} and E.~A. Cornell\cite{qpdNIST}}
\address{JILA, National Institute of Standards and Technology and Department of Physics, \\
University of Colorado, Boulder, Colorado 80309-0440}
\date{\today}

\maketitle

\begin{abstract}
We observe counter-intuitive spin segregation in an inhomogeneous
sample of ultra-cold, non-condensed Rubidium atoms in a magnetic
trap. We use spatially selective microwave spectroscopy to verify a
model that accounts for the differential forces on two internal spin
states. In any simple understanding of the cloud dynamics, the forces
are far too small to account for the dramatic transient spin
polarizations observed.  The underlying mechanism remains to be
elucidated.
\end{abstract}

 }
\par
Imagine a near-ideal gas consisting of a mixture of two distinct
molecular species. Apply a differential force that tends to pull the
two species apart. What determines whether the two species segregate,
or remain mixed? From a kinetics point of view, the question becomes
whether the segregating drift velocity, a function of the
differential force and the mean free path, is larger or smaller than
the effective velocity of remixing diffusion. The ``diffusive
velocity'' depends on the characteristic size of the sample. The
equilibrium point of view focusses instead on whether the energetic
benefit derived from separating in the differential potential
outweighs the corresponding entropic cost. Again, for a given
magnitude force, the ultimate outcome hinges on the size of the
sample.  Both points of view lead to the same quantitative criterion
-- is the differential potential across the sample large or small
compared to the thermal energy $k_BT$? As an example, a mixture of
gaseous oxygen and nitrogen in a room-sized box does not undergo
separation under the differential force of Earth's gravity, because
the gravitational potential of the differential mass (4 amu),
evaluated over a few meters, is very small compared to $k_BT$.
\par
In this paper, we study the segregation behavior of a gaseous
mixture. Our system consists of magnetically trapped ultra-cold
rubidium gas in two distinct spin states. These states can be
interconverted in the presence of an applied microwave magnetic
field, but in the absence of the applied drive the two states do not
interconvert and can be thought of as two distinct
fluids\cite{cornell1998a}. Inter-atomic interactions and differential
magnetic moments give rise to a small differential potential that
would tend to separate the two species, except that it is a thousand
times smaller than $k_BT$. In contrast to the above arguments we
observe that the initially homogenized sample nonetheless undergoes a
transient segregation.
\par
In the first part of this paper, we describe the use of spatially
selective Ramsey spectroscopy to characterize a differential
potential acting on the two spin states. In the second part, we
describe our observations of the atomic motion induced by the
differential potential over longer timescales. The atomic motions are
much too large to be described simply as arising from an acceleration
from the differential potential.
\par
The experimental apparatus includes a single magneto-optical
vapor-cell trap which precools a sample of $^{87}$Rb atoms for
transfer into a quadrupole magnetic trap. The electro-magnetic
quadrupole trap coils are mounted on a servo-controlled linear track
that mechanically moves the atoms to a UHV region where they are
transferred into a Ioffe-Pritchard magnetic trap for further cooling
via rf evaporation. Our Ioffe-Pritchard trap uses permanent magnets
to produce strong radial confinement. Electromagnetic coils generate
a stable and adjustable bias field and a weak confining potential in
the axial direction. The trapping frequencies are $\{7,230,230\}$ Hz.
Typical sample sizes are $10^6$ atoms at a temperature of 600 nK
(factor of 2 above the Bose-Einstein condensation temperature).
\par
The two hyperfine states of interest are $|1\rangle \equiv
|F=1,m_f=-1\rangle$ and $|2\rangle \equiv |F=2,m_f=1\rangle$. We
coherently couple the two hyperfine states via a two-photon
drive\cite{Matthews1998a}. The drive consists of a microwave photon
($\sim$ 6.8 GHz) and a rf photon (1-3 MHz). The sum frequency is
detuned 200 Hz from the transition, and the effective two-photon Rabi
frequency is 2.5 kHz. The drive is turned on for $\sim$ 100 $\mu$s to
transfer half of the atoms (initially all in the $|1\rangle$ state)
to the $|2\rangle$ state (a $\frac{\pi}{2}$ pulse). A second
$\frac{\pi}{2}$ pulse is applied after a variable delay . The final
population and spatial distribution in either spin state can be
separately determined using absorption imaging\cite{imaging}.
\par
\begin{figure} 
\begin{center}
\psfig{figure=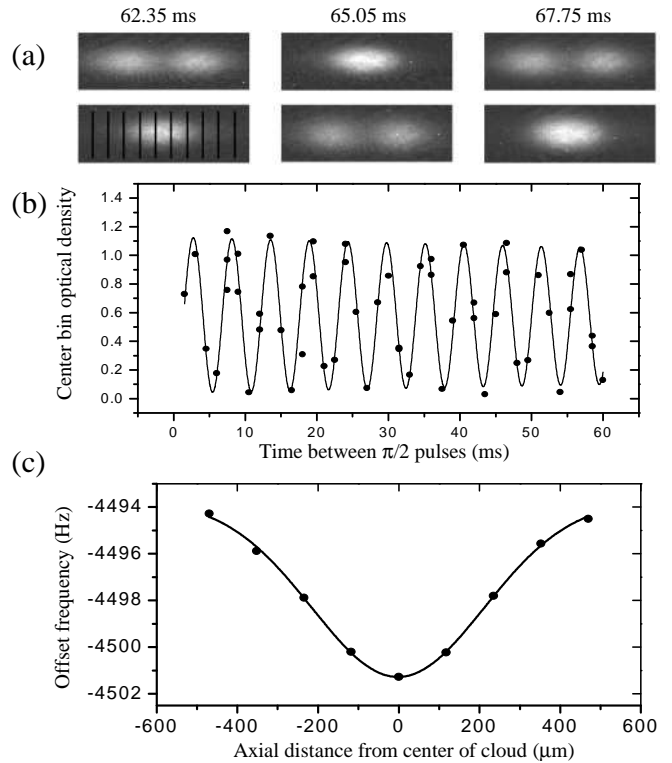,width=1\linewidth,clip=}
\end{center}
\caption {(a) Images of F=1 atoms (upper) and F=2 atoms (lower) after
two $\frac{\pi}{2}$ pulses. The delay between the pulses is listed at
the top. The edges of the cloud have become out of phase with the
center. The black lines show how the cloud is divided up along the
axial direction to measure the spatial dependence of the transition
frequency. Each image is 1 mm on the horizontal (axial) dimension.
(b) The center-bin $|1\rangle$ optical density as a function of time
between the two $\frac{\pi}{2}$ pulses. (c) The plot shows the
spatial dependence of the transition frequency in the axial
direction. Offset frequency is defined as the transition frequency -
6834682610.9 Hz. The solid line is a gaussian fit to the data. These
data were taken at $B_{bias}= 3.23$ G. }
 \label{composite}
\end{figure}

We determine the transition frequency between states $|1\rangle$ and
$|2\rangle$ by varying, in each realization of the experiment, the
delay time between the two $\frac{\pi}{2}$ pulses and measuring the
relative population in the $|1\rangle$ state. Typical delay times
range from 1-60 ms. The frequency of the resulting Ramsey fringes is
the difference between the atomic transition and the oscillator
frequency of our two-photon drive\cite{ramsey1956a}. In order to
study the spatial variation of the transition frequency between the
two states, we break up the images into 9 bins along the axial
direction and integrate the number of atoms in each bin (Fig.
\ref{composite}).
\par
There are two different mechanisms that shift the transition
frequency. The first, dominant in the limit of low atomic density, is
the differential Zeeman effect. The atoms are confined in the
parabolic minimum of the magnetic field and each spin state
individually experiences a Zeeman shift, which varies by 30 kHz from
the center of the cloud to the edge of the cloud \cite{size}. States
$|1\rangle$ and $|2\rangle$ were selected because the Zeeman shifts
cancel to great accuracy. The Breit-Rabi equation \cite{Breit}
predicts that the frequency splitting is first-order independent of
field at 3.23 G, increasing quadratically with magnetic field about
that minimum. The magnetic field in turn has a minimum value,
$B_{bias}$, at the center of the magnetic trap, and increases
quadratically with displacement along the trap axis. Thus for a
$B_{bias}$= 3.23 G, the differential Zeeman shift across the cloud is
quartic in axial position; the quartic shift from trap center to
sample edge is 0.7 Hz. For $|B_{bias} - 3.23 G|>$ 0.02 G, the
dominant field dependence with position is quadratic. For instance,
for $B_{bias}$ = 3G, the frequency shift from cloud center to edge is
-7 Hz.
\par
The second frequency shift is due to the mean-field energy that
arises from the self-interaction of the atoms. The two states have
slightly different s-wave scattering lengths and therefore different
interaction energies. The shift of the transition energy is given by

\begin{equation}
\Delta E_{mf}=\frac{4 \pi \hbar^2}{m}[ 2n_2 a_{22}-2n_1 a_{11}+
2(n_1-n_2) a_{12} ] \label{denshift} \label{meanequn}
\end{equation}
where $a_{ij}$ is the s-wave scattering length between states
$|i\rangle$ and $|j\rangle$, $n_i$ is the density in state
$|i\rangle$ and the prefactors of 2 account for the enhancement of
s-wave collisions for non-condensed bosons. For $^{87}$Rb the s-wave
scattering lengths are $a_{11}=100.9 a_0$, $a_{22}=95.6 a_0$, and
$a_{12}=98.2 a_0$\cite {heinzenpc} where $a_0$ is the Bohr radius.
The mean-field shift has very little dependence on relative
populations in the two states because $a_{12}\simeq
(a_{11}+a_{22})/2$. After a $\frac{\pi}{2}$ pulse, $n_1\approx n_2
\approx n/2$, and to sufficient accuracy $\Delta E_{mf} =
(a_{22}-a_{11})4 \pi \hbar n/m$.
\par
We measure the mean-field shift spectroscopically. The sample density
is inhomogeneous, with a gaussian profile in each dimension. The
radial oscillation period is short compared to the duration of the
Ramsey measurement and to the mean collision time ($\sim$ 10 ms).
Thus the random thermal motion of the atoms effectively averages over
the radial density inhomogeneity. The axial oscillation period, on
the other hand, is long compared to the measurement duration and to
the collision time. To a reasonable approximation one can treat the
atoms as ``frozen'' in the axial direction.  To further reduce the
effect of axial density inhomogeneity, we use the data only from the
center bin, a region of relative uniformity. We create clouds with
various peak densities and measure the transition frequency of the
center bin. The slope of the frequency vs. density fit $(-0.52 \times
10^{12}$ Hz/cm$^3$) is more negative than a prediction based on
photo-associative data by 20\% \cite{heinzenpc}. However, it is
entirely possible that our analysis of our cloud images could have
systematically underestimated the cloud density by as much as 20\%.
More troubling is that the zero-density extrapolation, at
6834678118.0(7)Hz at 3.23 G, is too high by 4.4(7) Hz.  We are unable
to account for this systematic offset.
\par
For a particular peak density we measure the axial spatial variation
in frequency by binning the cloud into several sections and
determining a frequency for each bin. This measurement is done at
$B_{bias}= 3.23$ G to minimize the magnetic contribution to the
inhomogeneity. The transition frequency vs. axial position along the
cloud is shown in Fig. \ref{composite}c.

\par
\begin{figure}  
\begin{center}
\psfig{figure=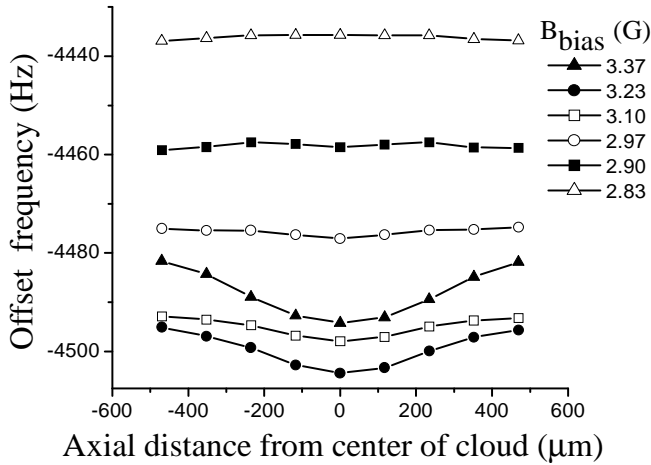,width=1\linewidth,clip=}
\end{center}
\caption {The transition frequencies at different positions along the
cloud are plotted as a function of $B_{bias}$. The cloud has a peak
density of 4.0 $\times 10^{13}$cm$^{-3}$. Offset frequency is defined
as the transition frequency - 6834682610.9 Hz. The frequency shift at
3.23 G (negligible magnetic contribution to the inhomogeneity) has a
gaussian shape. At $B_{bias}=3.37$ G the magnetic shift and the
mean-field shift are in the same direction and therefore the
inhomogeneity is across the cloud is increased. When the cloud is in
a field of 2.83 G inhomogeneity due to the mean-field shift is almost
completely cancelled by the magnetic shift. Data points are connected
to guide the eye.}
 \label{cancellation}
\end{figure}

We can change the frequency shift across the cloud and nearly cancel
both of the frequency shifts by operating at different a value of
$B_{bias}$. The cancellation of the mean-field and magnetic frequency
shifts can not be perfect because the cloud has a gaussian density
profile and the magnetic field shift is essentially parabolic across
the cloud. However, for a cloud with $1\times 10^6$ atoms at a
temperature of 500 nK there will be less than 1 Hz shift across the
cloud at $B_{bias}$= 2.83 G (Fig. \ref{cancellation}).

\par
Conversely, if we operate at $B_{bias} > 3.23$ G, the curvature of
the magnetic contribution to the frequency inhomogeneity adds with
the same sign to that from the density contribution, and we see
enhanced total frequency shift across the cloud (Fig.
\ref{cancellation}). Near the center of the cloud, the frequency
shift is parabolic with displacement from the center, and the
curvature of the differential potential can be conveniently
characterized by

\begin{equation}
\nu_{diff} = 1/2 \pi \sqrt{ h/m \times \langle d^2
\nu_{12}/{dz^2}\rangle}
\end{equation}
where m is the mass of $^{87}$Rb and $\nu_{12}$ is the transition
frequency\cite{dndz}. $\nu_{diff}$ may be thought of as the axial
motional frequency of a hypothetical test atom, in state $|2\rangle$,
under the influence of only the \emph{differential} magnetic and
mean-field forces. For the data shown in Fig. \ref{cancellation},
measured $\nu_{diff}$ ranges from 0.15 Hz for $B_{bias} = 3.37$ G,
down through 0.1\textit{i} Hz for $B_{bias} = 2.83$ G.
\par
We have used spectroscopic methods to characterize the spatial
dependence of the differential potential experienced by states
$|1\rangle$ and $|2\rangle$ in our trap. The data are quantitatively
accounted for by a simple model involving magnetic and mean-field
potentials. We now describe the observed mechanical effects that the
gradients of these potentials have on the distribution of the
relative atomic densities of states $|1\rangle$ and $|2\rangle$.
\par
In the twin-pulse spectroscopy described so far, the first pulse sets
up a coherent superposition of the two states; the delay time between
the two pulses allows the relative internal phase, $\phi$, in the
coherent superposition $(|1\rangle + e^{i \phi}|2\rangle )$ to evolve
under the inhomogeneous potential, and the second pulse then converts
the spatial pattern in $\phi$ into a pattern of relative density for
ready imaging. In the experiments described below, the second
$\frac{\pi}{2}$ pulse is omitted, and the delay between the first
pulse and the density imaging is increased to allow time for the
atoms to move under the influence of the developing spatial gradient
in $\phi$. The patterns in the density distribution which we image
must result from the \textit{physical displacement} of atoms during
the period after the $\frac{\pi}{2}$ pulse \cite{spinflip}.
\par
\begin{figure} 
\begin{center}
\psfig{figure=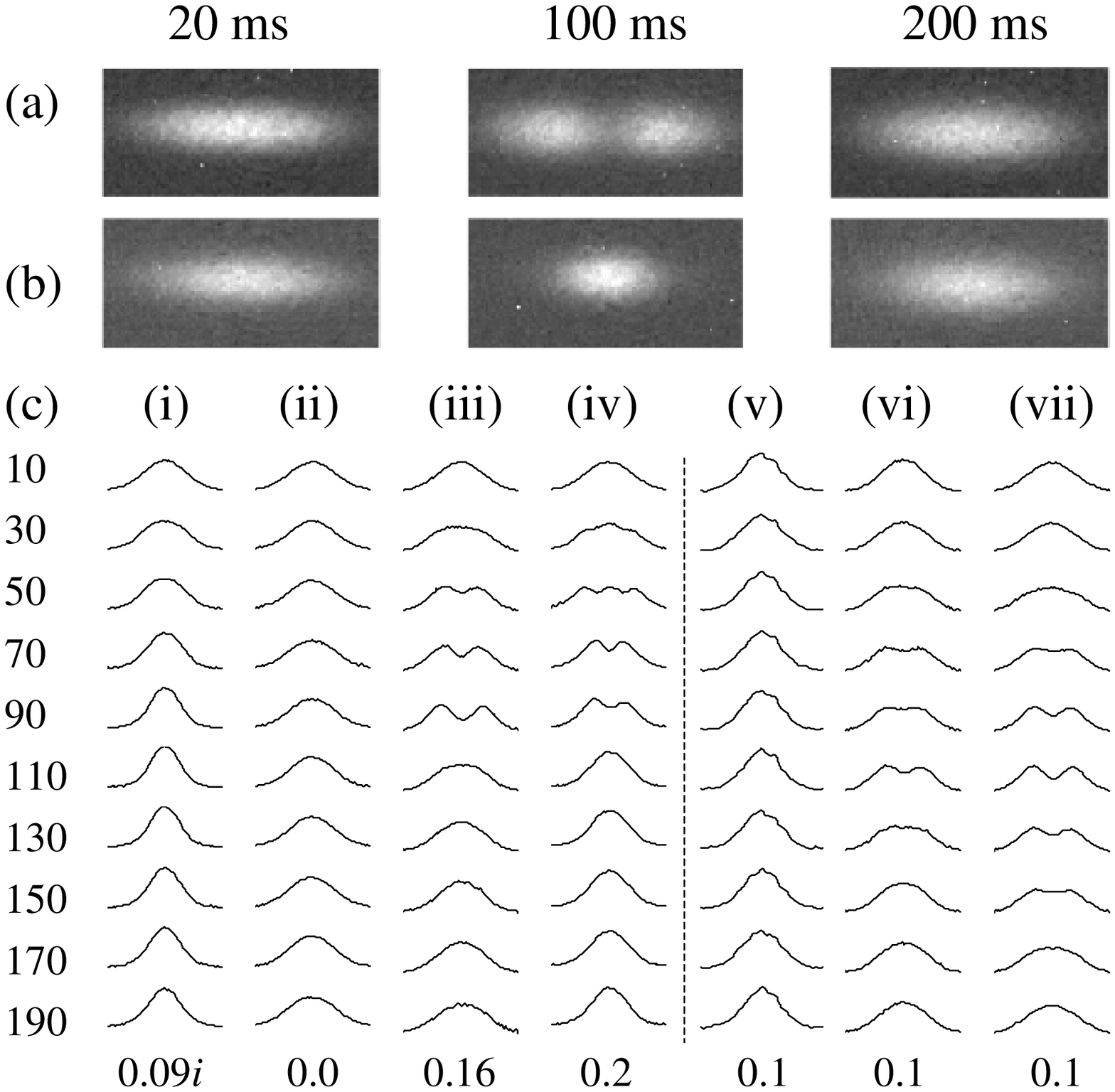,width=1\linewidth,clip=}
\end{center}
\caption { (a) Spin segregation appears in the spatial distribution
of atoms in the $|1\rangle$ state at various times after the
$\frac{\pi}{2}$ pulse. $|1\rangle$ state atoms move outward, then
relax back to their initial condition. Initial cloud conditions are
peak total density $1.8 \times 10^{13} $cm$^{-3}$, T=630 nK,
$\nu_{diff} = 0.1$ Hz. (b) Same as for (a), but showing the
$|2\rangle$ state distribution. $|1\rangle$ and $|2\rangle$-state
motions are complementary, and leave the total density profile
unchanged. (c) $|1\rangle$- state axial density profile imaged at
times after the $\frac{\pi}{2}$ pulse, listed on the left column .
For all columns, $T=600$ nK. $\nu_{diff}$ values are listed at the
bottom of each column. The first column of cross sections has an
imaginary $\nu_{diff}$ corresponding to negative curvature of the
frequency. Initial density (in $10^{13}$cm$^{-3}$) and $B_{bias}$ (in
Gauss) are for each column (i)1.82, 2.99 (ii)1.82, 3.12 (iii)1.82,
3.52 (iv)1.82, 3.79 (v)0.254, 3.35 (vi)1.16, 3.31 (vii)2.05, 3.27
Columns (i)-(iv) are at fixed density and progressively higher
$\nu_{diff}$, showing turn-on of effect with increasing $\nu_{diff}$.
Columns (v)-(vii) are at fixed $\nu_{diff}$ and at progressively
higher density, showing turn-on of segregation effect with density.}
 \label{separation}
\end{figure}

The results of this class of experiment are summarized in Fig.
\ref{separation}. Starting around 50 ms after the $\frac{\pi}{2}$
pulse, the atoms in the $|1 \rangle$ state are observed to move
outward along the trap axis, while the atoms in the $|2\rangle$ state
move inward. Maximum separation is realized at about 100 ms, then the
two clouds relax back to their original, well-overlapped distribution
by around 200 ms. No further evolution is observed. The spin
segregation is dependent on the curvature of the differential
potential. For $\nu_{diff} < 0.05$ Hz (Fig. \ref{separation},(ii)),
no separation is observed above the imaging noise. For $\nu_{diff}>
0.18$ Hz (Fig. \ref{separation},(iv)), we observe what we call
``higher-order'' effects, with the relative spin-density developing
multiple stripes.

\par
We find the segregation to be startling. While the mean-field and
magnetic forces can combine to provide a small buoyant force that
tends to push the $|1\rangle$ atoms out towards the lower density
regions of the clouds, the magnitude of the observed effect is
entirely disproportionate.  For $\nu_{diff}=0.1$ Hz, for instance,
the total frequency shift $\Delta \nu_{12}$  from center to edge of
cloud is 6 Hz. This should be compared to $k_BT/h \sim$ 10 kHz. The
criterion for separation discussed in the introduction is thus
clearly not satisfied. This objection would be surmounted if the
initially well-homogenized and non-condensed populations in each
spin-state nonetheless moved collectively (the relevant inequality
would then be $N_{atoms} (\Delta \nu_{12} h) \gg kT$ rather than
$\Delta \nu_{12} h \ll kT$)\cite{bec}. One would still need to
account, however, for the \emph{rate} of segregation. If the atoms
move ballistically in the differential potential, the minimum time
for the peak segregation to develop should be at least $1/(4
\nu_{diff})$\cite{bec}. But the observed segregation times are at
least 10 times faster.
\par
As seen in Fig. \ref{separation}c, to induce segregation one needs
not only a sufficiently large value of $\nu_{diff}$ but independently
one needs adequate density.  It is possibly significant that the
segregation seems to disappear when the density gets so low that the
cloud is no longer collisionally thick in the axial direction. While
the observed segregation rate is fast compared to 4 $\nu_{diff}$, it
is slow compared to $4 \nu_z$, where $\nu_z$ is the axial trap
frequency. That is, the random thermal speed of the atoms, if it were
selectively directed, is more than adequate to move the atoms the
required distance.  We speculate that as the internal phase $\phi$
develops a spatial gradient, it somehow acts as a ``switch'' to
control velocity-changing collisions such that atoms in the
$|2\rangle$ state are preferentially scattered inward and $|1\rangle$
atoms outward.
\par
Finally, the observation of non-diffusive evolution of spin
populations compels a comparison with the spin-wave phenomenon
observed \cite{spinwaveexp} and discussed extensively
\cite{spinwavetheory} in the 1980s. We understand the earlier work to
be an effect of the {\it transverse} spin. The present work describes
evolution of the relative population of energy eigenstates, a {\it
longitudinal} spin effect. Additional study may lead to connections
between the two spin effects.
\par
We acknowledge useful conversations with the other members of the
JILA BEC collaboration and with Kurt Gibble.  This work is supported
by the NSF and by NIST.


%


\begin{thebibliography}{1}
\bibitem{imaging}
We image the atoms in the $|1\rangle$ state by transferring them to
the $|F=2,m_f=-2\rangle$ state using a microwave frequency adiabatic
rapid passage. The antitrapped $|F=2,m_f=-2\rangle$ atoms expand
rapidly in the radial direction for 3-5 ms. We then probe the atoms
on the $|F=2,m_f=-2\rangle\rightarrow|F'=3,m'_{f}=-3\rangle$
transition. To image the atoms in the $|2\rangle$ state, we apply a
$\pi$ pulse, converting the atoms in the $|2\rangle$ state to the
$|1\rangle$ state and vice versa, and immediately proceed as with
state $|1\rangle$.
\bibitem{size}
We arbitrarily define the ``edges'' of the cloud to correspond to
$\pm 1.5 \times HWHM.$
\bibitem{Breit}
From the Breit-Rabi equation we expand the relative Zeeman shift
about the minimum: $\nu=$ 6834678113.59(2) Hz + 431.35957(9)Hz/G$^2$
$\times$ (B-3.228917(3) G)$^2$.  These values and associated
uncertainties are derived from the zero-field hyperfine splitting
6834682610.90434(3) \cite{salomon} and the nuclear g-factor
0.0009951414(10) \cite{Ksplitting}.
\bibitem{dndz}
We use a density weighted frequency curvature defined by $\langle
\frac{d^2\nu_{12}}{dz^2} \rangle =
\int^{\infty}_{-\infty}n(z)\frac{d^2 \nu_{12}}{dz^2}(z)dz/\int
^{\infty}_{-\infty} n(z)dz$ where n(z) is the axial gaussian density
profile.
\end{thebibliography}
\end{document}